\documentclass[twocolumn,preprintnumbers,amsmath,amssymb]{revtex4}
\usepackage{amsmath,amssymb,graphics,epsfig,subfigure}
\usepackage{color}

\def\be{\begin{equation}}
\def\ee{\end{equation}}
\def\bea{\begin{eqnarray}}
\def\eea{\end{eqnarray}}

\begin{document}

\thispagestyle{empty}

\begin{center}

\title{Ruppeiner curvature along a Renormalization group flow}

\date{\today}
\author{Pavan Kumar Yerra\footnote{E-mail: pk11@iitbbs.ac.in} and
        Chandrasekhar Bhamidipati \footnote{E-mail: chandrasekhar@iitbbs.ac.in} }
 \affiliation{School of Basic Sciences,
Indian Institute of Technology Bhubaneswar, Bhubaneswar 752050, India}

\begin{abstract}
We report on an investigation of Ruppeiner curvature $R_U$ for $(d+1)$-dimensional  hyperbolic black holes in anti-de Sitter spacetimes and its connection with Renormalization Group (RG) flows in the dual d-dimensional conformal field theories (CFTs) in Minkowski spacetimes. A repulsive type interaction among microstructures is found, which is weaker for positive mass black holes and grows stronger for negative mass black holes at low temperatures. In particular, we show that $R_U$ evaluated along a zero mass curve is a universal constant, depending only on the dimension of space-time.  The extremal black holes are pointed out to have a positive Ruppeiner curvature irrespective of their horizon topology.

\end{abstract}

%%\pacs{}

\maketitle
\end{center}
%%%%%%%%%%%%%%%%%%%%%%%

\section{Introduction}

\noindent
Unraveling the nature of microscopic degrees of freedom of black holes has been a major goal of quantum gravity ever since the formulation of laws of black hole thermodynamics\cite{Bekenstein:1973ur,Bekenstein:1974ax,Hawking:1974sw,Hawking:1976de}, where the area and surface gravity were identified respectively with entropy and temperature. These works showed that black holes are not just interesting gravitational objects but also rather special thermodynamic systems, where their entropy $S$ is proportional to area $A$, instead of volume. \\

\noindent
Following Boltzmann's dictum, ``If you can heat it, it has microstructure'', black holes with non zero Hawking temperature are expected to have microscopic structure. For day to day fluid systems where a clear picture of microscopic molecular structure is available, the bottom up approach of statistical mechanics can be used to obtain information about macroscopic thermodynamic quantities. 
An indispensable tool of statistical mechanics used here is the Boltzmann's formula for entropy given as:
\begin{equation} \label{entropy}
S = k_B \, \ln \, \Omega\, ,
\end{equation}
where $k_B$ is the Boltzmann constant and $\Omega$ is the number of microstates of the respective thermodynamic system. In this direction, several different methods have been employed for black holes, including attempts at the computation of entropy via microscopic counting of states in special extremal black holes and general situations~\cite{Vafa,Maldacena,Callan,Horowitz,Emparan,Lunin,Mathur}. Deeper inroads have been made in understanding microscopic aspects in a holographic setting\cite{Maldacena:1997re, Gubser:1998bc, Witten:1998qj, Witten:1998zw}. The entanglement entropy\cite{Ryu:2006bv,Ryu:2006ef} of the ground state of a conformal field theory (CFT) reduced to a region bounded by a ball, is shown to be computed as the thermal entropy of a $(d+1)$-dimensional asymptotically anti-de Sitter (AdS) space-time in a special hyperbolic slicing\cite{Emparan:1999pm,Emparan:1999gf,Casini:2011kv}. In fact,  the universal contribution to entanglement entropy of a sphere\cite{Casini:2011kv}:
\begin{equation} \label{ad}
a^*_d = \frac{\pi^{d/2-1}\, L^{d-1}}{8\, G\, \Gamma(d/2) }\, ,
\end{equation}
is conjectured to be a decreasing quantity along renormalization group (RG) flow\cite{Myers:2010xs,Myers:2010tj}. Here, $L$ is the radius of AdS and $G$ is the Newton's constant. Analysis of the above connection shows that $a^*_d$ is a generalized central charge keeping track of the number of degrees of freedom of CFTs, a higher dimensional analogue of central charge in two dimensions. In even dimensions, the above conjecture coincides with that proposed by Cardy~\cite{Cardy:1988cwa} and is in agreement with proof of c-theorem discussed by Casini {\it et.~al.} in~\cite{Casini:2004bw,Casini:2012ei,Casini:2017vbe}. \\

\noindent 
Our aim is to employ a top down macroscopic approach of Ruppeiner geometry (an inverse process of statistical mechanics)~\cite{Ruppeiner} for hyperbolic black holes in AdS, to make possible predictions on the nature of microscopic degrees of freedom. The idea of Ruppeiner geometry stems from studying thermodynamic fluctuation theory, whose starting point is the inversion of the formula in eqn. (\ref{entropy}), giving
\begin{equation}\label{omega}
\Omega = e^{\frac{S}{k_B}}\, .
\end{equation}
Consider a large thermodynamic system $I_0$ in equilibrium, which contains in it a sub-system $I$, with the later allowed to have, two independent fluctuating variables, $x^i$ where $i=1,2$. 
The probability $P(x^1,x^2)$ of finding the state of the system between $(x^1,x^2)$ and $(x^1 + dx^1,x^2+dx^2)$ is proportional to number of microstates $\Omega$ from eqn. (\ref{omega}). The statement of second law of thermodynamics is that the pair $(x^1,x^2)$  takes on those values which maximize the entropy $S=S_{\text{max}}$ or in other words, the pair describes fluctuations around this maximum. Expanding the entropy around this maximum up to second order shows that~\cite{Ruppeiner:1995zz}:
\begin{equation}
P(x^1,x^2) \propto e^{-\frac{1}{2}\, \Delta l^2} \, ,
\end{equation}
with
\begin{equation}\label{distance}
\Delta l^2 \, = -\frac{1}{k_B}\, \frac{\partial^2 S}{\partial x^i\partial x^j}\, \Delta x^i \Delta x^j,
\end{equation}
denoting the line element for measuring thermodynamic distance between two neighboring fluctuation states. The more probable a fluctuation between two states, the shorter the distance between them. From study of thermodynamic curvature $R$ following from the line element in eqn. (\ref{distance}) for various fluid systems, such as, ideal and van der Waal (vdW) fluids, Ising models, Bose/Fermi systems and quantum gases, the empirical understanding emerging is that: A positive (negative) value of $R$ indicates the dominance of repulsive (attractive) type interactions.
A larger positive (negative) value of $R$ means the system is more (less) stable, which is indicative of the respective stability of Fermi (Bose) system~\cite{Ruppeiner}. Vanishing of $R$ means balance of attractive and repulsive interactions, i.e., a non-interacting situation. \\

\noindent
In the current setting where entropy is clearly the quantity to start with, the line element can be constructed as follows. Taking the internal energy $U=U(S,V)$ as the starting point, the first law reads $dU = T dS - P dV$, which can be inverted to give:
\begin{equation}
dS = \frac{1}{T} dU + \frac{P}{T} dV \, .
\end{equation}
In this plane, taking the fluctuating variables to be  $(x^1=T,x^2=V)$ the line element can be shown to be~\cite{Landau,Wei:2019uqg,Wei:2019yvs}:
\begin{equation} \label{metricU}
dl^2_{U} = \frac{1}{T} \left(\frac{\partial S}{\partial T} \right) \Bigr|_V dT^2 - \frac{1}{T} \left(\frac{\partial P}{\partial V} \right) \Bigr|_T dV^2\, .
\end{equation}
Note that the line element above in eqn. (\ref{metricU}) is quite generic and valid for any thermodynamic system (not necessarily a black hole), as specific expressions for thermodynamic quantities are not used in obtaining it.  In other words, eqn. (\ref{metricU}) has to be evaluated on a case to case basis and then used to compute the corresponding curvature $R$. The said line element was in fact used in earlier studies, to investigate the microstructures of van der Waals system~\cite{Wei:2019uqg,Wei:2019yvs}. Now, for the specific case of black holes in AdS, $V$ is the thermodynamic volume conjugate to  pressure $P = - \Lambda/{8\pi G}$~\cite{Kastor:2009wy,Dolan:2010ha,Cvetic:2010jb}. Using expressions for thermodynamic quantities, the corresponding curvature $R$ can be obtained. For certain classes of examples, such as static black holes in AdS with spherical horizons, additional normalization of Ruppeiner curvature may be required to extract the information about microstructures~\cite{Wei:2019uqg,Wei:2019yvs}. This will in fact also be the case in the current system of hyperbolic black holes, which will be elaborated further in section-(\ref{two}).\\

\noindent
Ruppeiner geometry has given invaluable insights~\cite{Wei:2019uqg,Wei:2019yvs,Xu:2019gqm,Ghosh:2019pwy}, in particular for black holes in AdS with spherical horizons. It was shown that the phase transitions~\cite{Chamblin:1999tk,Chamblin:1999hg,Cai:1998ep,Kubiznak:2012wp} might be explained as coming from interactions of microstructures being effectively attraction or repulsion dominated~\cite{Wei:2015iwa,Wei:2019uqg,Wei:2019yvs}, including the emergence of novel universal constants at the Hawking-Page transition~\cite{Wei:2020kra}. Our aim in this work is to explore the case of hyperbolic black holes in AdS whose phase structure we study in section-(\ref{two}) from Ruppeiner geometry point of view. In section-(\ref{three}), we present a discussion on the extremal and large dimension limit of thermodynamic curvature with interesting consequences.\\

\noindent
\section{Thermodynamic curvature of hyperbolic black holes} \label{two}
The metric for hyperbolic black holes in AdS in $(d+1)$ spacetime dimensions is\cite{Birmingham:1998nr}:
\begin{equation}
	ds^2=- f(\rho)d\tau'^2+\frac{d\rho^2}{f(\rho)}+\rho^2\left(du^2+\sinh^2(u)d\Omega^2_{d-2}\right)\, ,
\end{equation}
with the lapse function
\begin{equation}
	f(\rho)=\frac{\rho^2}{L^2}-\frac{\mu}{\rho^{d-2}}-1\ .
\end{equation}
$\mu$ is related to the (ADM) mass of the black hole:
\begin{eqnarray}\label{eq:mass}
 M(\rho_+,L)&=& \left(\frac{(d-1)w_{d-1}}{16\pi G}\right)\mu \nonumber \\ 
&=& \left(\frac{(d-1)w_{d-1}}{16\pi G}\right) 
	\rho_+^{d-2}\left[\left(\frac{\rho_+}{L}\right)^2-1\right]\,,
\end{eqnarray}
where $w_{d-1}$ stands for the volume of the hyperbolic space of unit radius and $\rho=\rho_+$ is the horizon radius. General properties of these black holes in arbitrary dimensions were discussed in \cite{Mann:1996gj,Vanzo:1997gw,Brill:1997mf,Emparan:1998he,Emparan:1999pm,Emparan:1999gf,Cai:2004pz}.
The black hole  temperature $T$, thermodynamic volume $V$, entropy $S$ and pressure $P$ for $d \geq 3$ are given by:
\begin{eqnarray} 	
	\label{eq:the-temperature}
	T&=&\frac{(d-2)}{4\pi}\frac{1}{\rho_+}
	\left[\frac{16\pi G }{(d-1)(d-2)}\rho_+^2p-1\right]\ ,
	\\
	\label{eq:the-volume}
	V&=&\frac{w_{d-1}\rho_+^{d}}{d} ,
	\\
	\label{eq:the-entropy}
	S&=&\frac{w_{d-1}\rho_+^{d-1}}{4G}\ ,
	\\ \label{eq:pressure}
	p&=&-\frac{\Lambda}{8\pi G}=
\left(\frac{d(d-1)}{16\pi G}\right)\frac{1}{L^2}\, .
\end{eqnarray}
The equation of state can be written from~\eqref{eq:the-temperature} as~\cite{Johnson:2018amj}:
\begin{equation}
\label{eq: eqofst}
p =\frac{( d-1)}{16\pi G} \bigg(\frac{Vd}{\omega_{d-1}}\bigg)^{\frac{-2}{d}}\bigg( 4 \pi T \Big(\frac{Vd}{\omega_{d-1}}\Big)^{\frac{1}{d}} + d-2\bigg),
\end{equation}
%\begin{figure}[h]
\begin{figure}
%	{
%		\centering
		\includegraphics[width=0.4\textwidth]{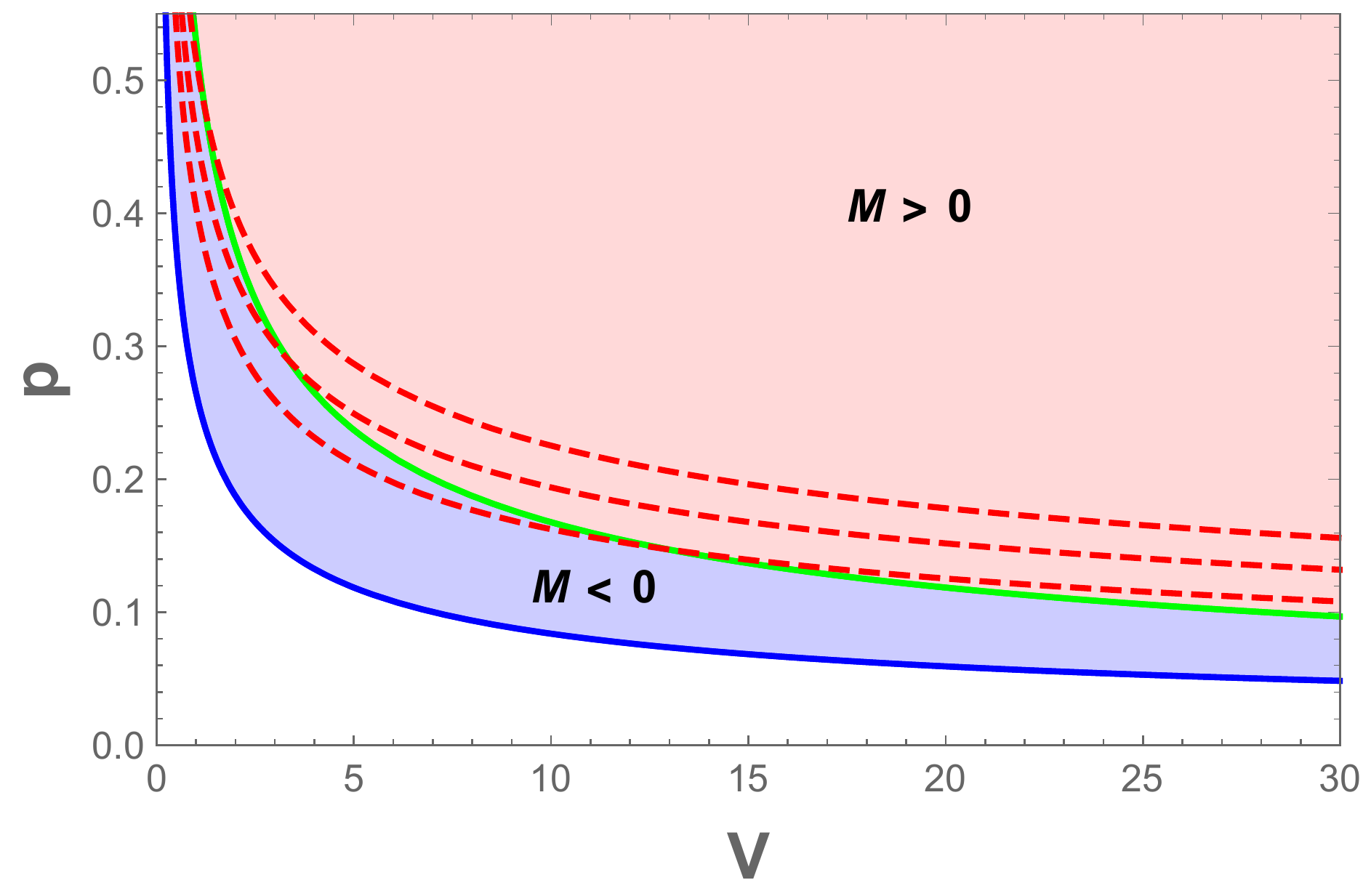}
		\caption{\label{fig:pv_plot} pV plot for 5-dimensional case:  Isotherms are shown in dashed red color whose temperature decreases from top to bottom $ (T=0.225, 0.175, 0.125)$, $T=0$ isotherm is shown with blue color, while $M=0$ curve  is shown with green color. The region above the  $M =0$ curve has $M  > 0$, while the region below the $M =0$ curve (including the $T=0$ curve) has $M < 0$. Except the $T=0$ isotherm, every isotherm crosses  the  $M=0$ curve.}
%	}
\end{figure}
and its behavior  for various isotherms can be seen from figure-\eqref{fig:pv_plot}, which shows that there are no phase transitions unlike the case of black holes with spherical horizons.
Since the entropy $S$ and thermodynamic volume  $V$ are not independent, we note that the heat capacity at constant volume, $C_V = 0$, while, the  heat  capacity at constant pressure $C_p$ is positive and computed to be:
\begin{equation}
\label{eq:cp}
C_p = \frac{(d-1)\omega_{d-1}}{4G}\rho_+^{d-1} \frac{(16\pi Gp \rho_{+}^2 -2 +3d -d^2)}{(16\pi Gp \rho_{+}^2 +2 -3d +d^2)}.
\end{equation}
Hyperbolic black holes in AdS have some unique properties in that their phase structure is dominated by a black hole at any temperature. Apart from a general mass $M>0$ solution, there is in particular a $M=0$ solution where the solution is isometric to AdS, but covers only a portion of the full geometry with a Rindler type horizon. It should be mentioned that, although this solution has a temperature with non-vanishing area (in $d=4)$, it is not a proper black hole due to absence of singularities. There is also a zero temperature solution (with $M<0$) which is different from the solution isometric to AdS and admits an extremal limit. \\

\begin{figure}
%\begin{wrapfigure}{l}[10]{0.4\textwidth}
		\includegraphics[width=0.4\textwidth]{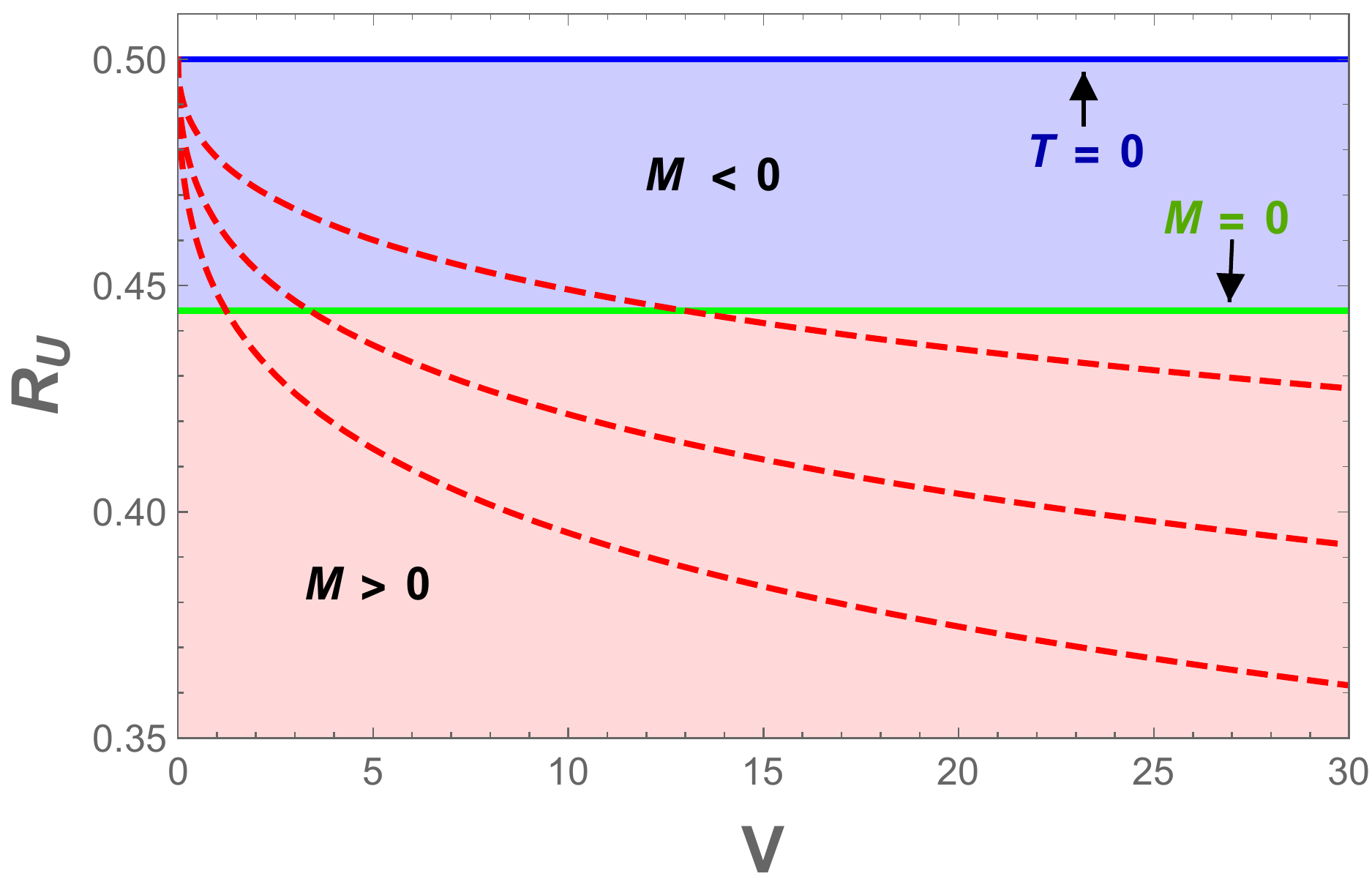}
		\caption{\label{fig:RUplot} The scalar curvature $R_\text{U}$ as a function of the volume V
for  5-dimensions and $T=0.225, 0.175$, and $0.125$ from bottom to top (dashed red color). $R_\text{U}=\frac{1}{2}$  (Thick blue line at the top) when $T=0 $, and  $R_\text{U}=4/9$ (green line in the middle) when $M=0$. The shaded regions correspond to $M<0$ (blue) and $M>0$ (pink).}
\end{figure}

\begin{figure}
		\includegraphics[width=0.3\textwidth]{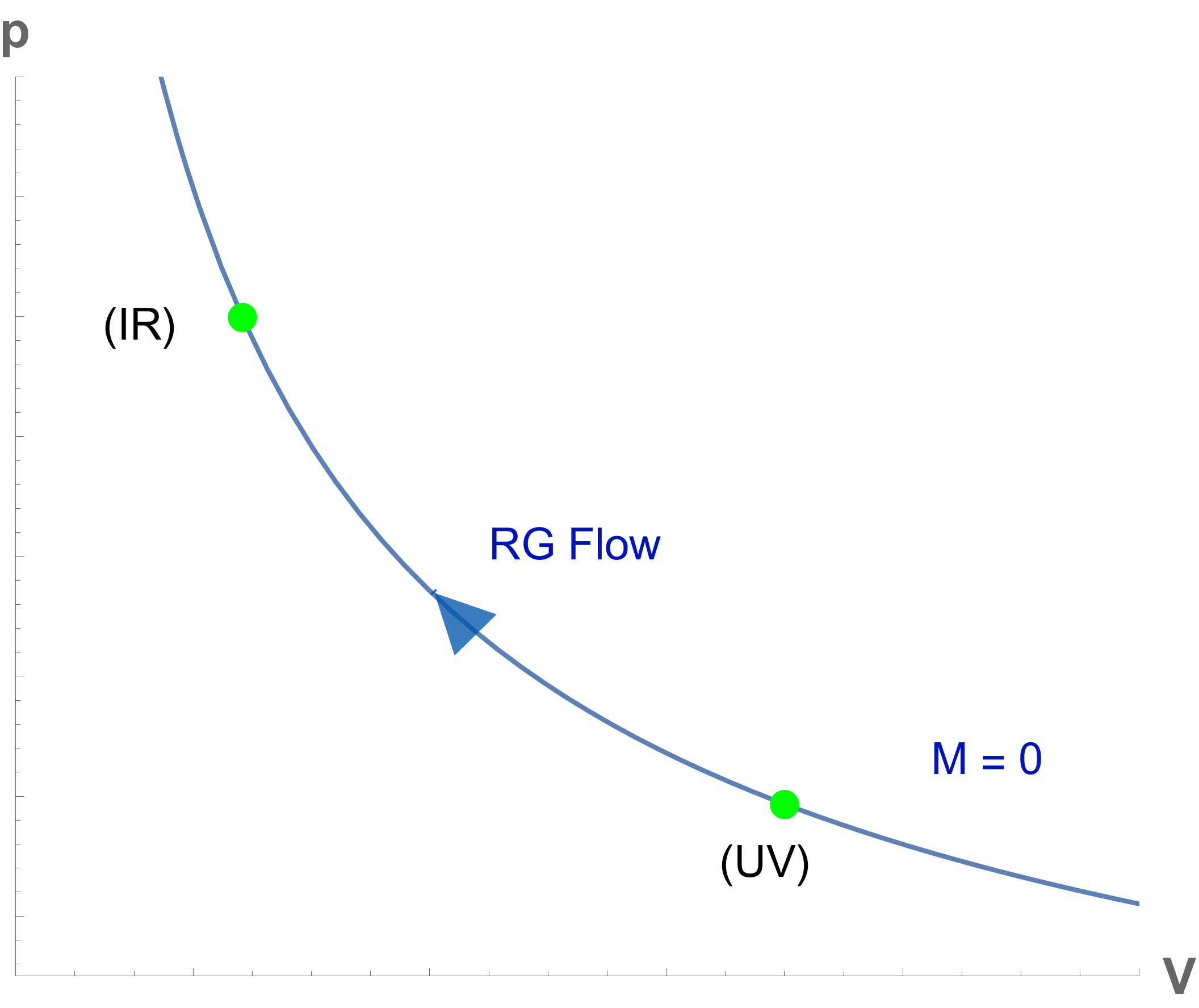}
		\caption{\label{fig:rg_flow}The zero mass curve representing the RG flow.}
	\end{figure}

\noindent
Along the zero mass curve in the $(p,V)$ plane, pressure, temperature and $C_p$ simplify as~\cite{Johnson:2018amj}:
\begin{equation}\label{eq:m=0eqofst}
	p^{ \ (M=0)}=\frac{\kappa}{V^{2/d}}\ , \qquad T^{ \ (M=0)} = \frac{1}{2\pi L},
\end{equation}
\begin{equation}
\label{eq:m=0Cp}
C_p^{ \ (M=0)} = \frac{4^{-d} \big( \frac{d(d-1)}{Gp} \big)^{\frac{d-1}{2}} \pi^{\frac{1-d}{2}}\omega_{d-1}}{G}\, ,
\end{equation}
where
\begin{equation}
	\kappa=\frac{d(d-1)}{16\pi G}\left(\frac{w_{d-1}}{d}\right)^{2/d}\ .
\end{equation}
\noindent
Now, to study the effect of temperature-volume fluctuations on microstructure interactions in hyperbolic black holes of general mass, we use eqns. (\ref{eq:the-temperature})-(\ref{eq:pressure}) to compute the line element in eqn. (\ref{metricU}). An immediate problem one encounters is that $\left(\frac{\partial S}{\partial T} \right)_V= C_V/T$ is zero, rendering the metric non-invertible. This can be dealt with by treating the vanishing of heat capacity $C_V$ as the $k_B \rightarrow 0^{+}$ limit. That is, $C_V$ is treated to be a small positive quantity, whose value is infinitely close to zero as proposed in~\cite{Wei:2019uqg,Wei:2019yvs}. With this, the curvature $R$ can be computed straightforwardly, and since $C_V$ comes out to be an overall multiplicative factor, one can define a new normalized curvature as $R_U = R \, C_V$~\cite{Wei:2019uqg,Wei:2019yvs}. Note that the empirical information about microstructure interactions contained in $R$ and $R_U$ is essentially the same, since they are related by a positive constant. $R_U$  is of course more practical to use for static black holes and it would be nice to know whether there is a deeper reason why this normalization procedure works. In what follows, we use $R_U$ to probe the microstructure of hyperbolic black holes in AdS and for the present case we obtain:
\begin{equation}\label{eq:RN}
R_\text{U} =  \frac{(d-2)}{2} \frac{\Big(4\pi T \big( \frac{Vd}{\omega_{d-1}}\big)^{\frac{1}{d}} + d-2 \Big)}{\Big(2\pi T \big( \frac{Vd}{\omega_{d-1}}\big)^{\frac{1}{d}} + d-2 \Big)^2}.
\end{equation}
\noindent
Assuming an appropriate regularization for $\omega_{d-1}$~\cite{Johnson:2018amj}, $R_{\text{U}}$ plotted in figure-(\ref{fig:RUplot}) is positive for any M and at any temperature, with the $M=0$ line seen to clearly separate the $M>0$ and $M<0$ black holes cases.  The curvature starts out at a higher value at smaller volumes, where the negative mass black holes dominate, and continues to decrease, with lowest values occurring for positive mass black holes at high volumes. $R_{\text{U}}$ does not have any divergences signifying that there are no phase transitions, matching well with traditional understanding that the dual CFT is always in a deconfined phase~\cite{Emparan:1999gf}. \\
%In fact, it is the $T=0$ extremal black hole with $M<0$ which has the highest positive value, signifying strongly correlated microstructure states, which .
\\
\noindent
Ruppeiner curvature computed along the zero mass curve gives an intriguing result
\begin{equation}\label{eq:m=0RN}
R^{M=0}_\text{U}  =  \frac{d(d-2)}{2(d-1)^2},
\end{equation}
which is positive and a universal constant (i.e., depending only on the number of spacetime dimensions).  The sign of $R^{M=0}_\text{U}$ empirically suggests that the microstructure interactions are predominantly of repulsive type and possibly strongly correlated.  This result should be contrasted with the case of Schwarzschild black holes in AdS with spherical horizons considered in\cite{Wei:2020kra}, where the curvature comes out to be a universal constant as well, but with negative sign. Further, from the observations in~\cite{Johnson:2018amj}, where it was argued that moving towards higher values of $p$ along the massless curve in figure-(\ref{fig:rg_flow}) corresponds to an RG flow~\cite{Girardello:1998pd,Distler:1998gb},
One might speculate the following. Volume fluctuations (a horizontal movement in figure-(\ref{fig:rg_flow})) correspond to perturbing the ground state of CFT. Thus, the CFT ground state is stable against small perturbations, the corresponding possible effect is that strength of interactions captured by $R^{M=0}_\text{U}$ should not change. 
It would of course be nice to get a precise understanding of the reason for constancy of $R^{M=0}_\text{U}$. \\

\noindent
\section{Discussion}\label{three}
Few remarks on the $T=0$ extremal limit are worth noting, as it has unusual properties of  large degeneracy with finite entropy (in five dimensions), despite having zero energy density, in tension with the third law of thermodynamics~\cite{Emparan:1999pm}. The weak coupling analysis suggests that the low temperature regime is dominated by spin-1/2 and spin-1 fields, although the strong coupling degrees of freedom are unclear~\cite{Emparan:1999pm}. Our analysis shows that at low temperatures $R_\text{U}$ is positive and taking maximum value of half. The sign and nature of Ruppeiner scalar~\cite{RupMay} empirically suggests that from thermodynamic fluctuation theory point of view, the overall collective behavior of strong coupling degrees of freedom at low temperatures is dominated by a repulsive type of interaction.  What is the surprising is that, for charged black holes in AdS with spherical horizons, there exist both attractive ($R_U <0$ regions) and repulsive ($R_U >0$ regions) interactions. These compete at various temperatures possibly resulting in phase transitions, but the $T=0$ behavior is universal, in that there are only repulsive type interactions where $R_U =1/2$ for charged spherical and neutral hyperbolic black holes (although the near horizon geometries are different), and $R_U =1/3$ for the vdW system. The above observations, possibly suggest that the microstructures of all extremal black holes irrespective of their horizon topology, are dominated by repulsive interactions with positive Ruppeiner curvature. It would be interesting to verify this in detail by studying the behavior of microstructures in more general systems, such as, rotating black hole spacetimes, and with hyperbolic topology in charged and higher derivative theories to test the above proposal using other possible approaches~\cite{Sen:2014aja}.\\

\noindent 
Another amusing fact related to the above discussion comes to light from taking the interesting large $d$ limit in General relativity~\cite{Emparan:2013moa,Bhattacharyya:2015dva}. The thermodynamic geometry for various black hole systems in higher dimensional cases have been studied in literature~\cite{Wei:2019yvs,Wei:2020kra,HosseiniMansoori:2020jrx,Wei:2021pql}.  However, the large d-limit of thermodynamic curvature has not been reported earlier, and in fact, a general study of thermodynamic geometry in this limit is interesting in its own right.  Now, in the context of hyperbolic black holes, the efficiency of holographic heat engines ($M>0$ case) approaches the maximum allowed Carnot efficiency, as the number of dimensions increases~\cite{Johnson:2018amj}. Interestingly, in the large $d$ limit, $R^{M=0}_\text{U}$ in eqn. (\ref{eq:m=0RN}) increases and finally approaches the maximum possible value of $1/2$, which is the same value obtained in the extremal limit discussed above, pointing towards flow of the system to a state where the interactions are strong and stable. Reason for $\lim_{d \rightarrow \infty}R^{M=0}_\text{U} = R^{T=0}_\text{U}$ is that, when $M =0$ we have  $\rho_+  = L$, while when $T=0$ we have $\rho_+ = L \sqrt{(d-2) /d} $, which goes over to the former in large $d$ limit. We now compare the above results with the large d behavior of Ruppeiner curvature for black holes with spherical horizons.  From the results in~\cite{Wei:2020kra}, $R_U$ at the Hawking-Page transition point is negative and diverges as the number of dimensions increases, pointing towards strong correlations and possibly an instability~\cite{Janyszek1990}. The precise consequences of these observations can be understood better by investigating the thermodynamic geometry of black holes in the large d limit (in general, and not necessarily at special points in the parameter space) and should be pursued in future.\\

\noindent
In summary, starting from the Boltzmann entropy formula, we constructed Ruppeiner geometry of hyperbolic black holes in AdS space-times in arbitrary dimensions.  Employing $(T,V)$ as fluctuating coordinates, we computed thermodynamic curvature $R_\text{U}$ and showed its usefulness in predicting the nature of black hole microstructures. Certain universal relations using Ruppeiner geometry were found earlier for black holes in AdS with spherical horizons  at the Hawking-Page transition point in~\cite{Wei:2020kra}. Black holes with spherical horizons are known to have both attractive and repulsive type interactions\cite{Wei:2019uqg}. On the other hand, we reported here the analysis for the AdS black holes with hyperbolic horizons which shows the presence of only repulsive interactions, quite different from the spherical horizon counterpart. In particular, novel universal features were found by exploring their nature along the zero mass curves. $R^{M=0}_\text{U}$ turned out to be a universal constant, unchanged along the RG flow curve and indicating that the interactions of microstructures are stable against  temperature and volume fluctuations. Further  connection of our results in the context of dual field theory is worth exploring. The interplay of thermodynamics and geometric aspects of black holes explored in this work and possible connections in the AdS/CFT context, should be studied further as they might help us understand the important physics of microscopic degrees of freedom.

\section*{Acknowledgements}
One of us (C.B.) would like to thank the DST (SERB), Government of India, for financial support through the Mathematical Research Impact Centric Support (MATRICS) grant no. MTR/2020/000135 and Aritra Ghosh for helpful discussions. We also thank the anonymous referees for several helpful suggestions which improved the manuscript.

%\bibliographystyle{apsrev4-1}

%\bibliography{rg}

\end{document}